\documentstyle[12pt]{article}

\begin{document}

\title{MULTISPHALERONS IN THE WEAK INTERACTIONS}
\vspace{1.5truecm}
\author{
{\bf Burkhard Kleihaus}\\
Fachbereich Physik, Universit\"at Oldenburg, Postfach 2503\\
D-26111 Oldenburg, Germany
\and
{\bf Jutta Kunz}\\
Fachbereich Physik, Universit\"at Oldenburg, Postfach 2503\\
D-26111 Oldenburg, Germany\\
and\\
Instituut voor Theoretische Fysica, Rijksuniversiteit te Utrecht\\
NL-3508 TA Utrecht, The Netherlands}

\vspace{1.5truecm}

\date{March 15, 1994}

\maketitle
\vspace{1.0truecm}

\begin{abstract}
We construct multisphaleron solutions in the weak interactions.
The multisphaleron solutions carry Chern-Simons charge $n/2$,
where $n$ is an integer.  The well-known sphaleron has
$n=1$ and is spherically symmetric for vanishing mixing angle.
In contrast the multisphalerons with $n>1$
are only axially symmetric, even for vanishing mixing angle.
The greater $n$, the stronger the energy density deforms.
While for small Higgs masses the energy of the multisphalerons is smaller
than $n$ times the energy of the sphaleron,
the energy of the multisphalerons is larger
than $n$ times the energy of the sphaleron
for large Higgs masses.
\end{abstract}
\vfill   \noindent Univ. Utrecht-Preprint THU-94/04
\vfill\eject

\section{Introduction}

In 1976 't Hooft [1] observed that
the standard model does not absolutely conserve
baryon and lepton number
due to the Adler-Bell-Jackiw anomaly.
In particular 't Hooft considered spontaneous
fermion number violation due to instanton transitions
between topologically inequivalent vacua.
Fermion number violating
tunnelling transitions at high energies
might become observable at future accelerators [2,3].

In 1983 Manton [4] considered
the possibility of fermion number
violation in the weak interactions
from another point of view.
He showed that there are non-contractible
loops in configuration space and
predicted the existence of a static, unstable solution
of the field equations,
a sphaleron [5], representing
the top of the energy barrier between
topologically distinct vacua.

It is this interpretation of the sphaleron solution,
which makes it physically relevant.
At finite temperature the energy barrier between
distinct vacua can be overcome
due to thermal fluctuations of the fields,
and vacuum to vacuum transitions can occur,
accompanied by a change of baryon and lepton number.
The rate for baryon number violating processes
is largely determined by a Boltzmann factor,
containing the height of the barrier at a given
temperature, and thus by the energy of the sphaleron [6-10].

In this letter we investigate the configuration space of
the Weinberg-Salam theory further and construct
multisphaleron solutions.
In contrast to the sphaleron [5],
which is spherically symmetric for vanishing mixing angle,
the multisphaleron solutions are only axially symmetric,
even for vanishing mixing angle.
The appropriate ansatz for the multisphalerons represents
a generalization of the
axially symmetric ansatz for the sphaleron at
finite mixing angle [11,12],
preserving the invariance under parity.

Like the multiinstantons the multisphalerons can be labelled
by an integer $n$,
which is associated with the topological charge of these solutions.
This integer $n$ appears in the ansatz for the multisphalerons,
representing a winding number with respect to the azimuthal angle
$\phi$. While $\phi$ covers the full trigonometric circle once,
the fields wind $n$ times around.
The ansatz is thus
analogous to the ansatz of Manton [13]
and Rebbi and Rossi [14]
for multimonopole solutions
of an SU(2) gauge theory with a Higgs triplet.

{}From the ansatz the energy functional and the equations of motion
for the fields are obtained.
Considering only vanishing mixing angle,
the resulting system of coupled non-linear partial
differential equations is solved numerically.
For $n=1$ spherical symmetry and the well known sphaleron [5]
are recovered.

In section 2 we briefly review the Weinberg-Salam lagrangian,
discuss the ansatz and present the resulting energy functional.
In section 3 we exhibit the multisphaleron
solutions with $n \le 5$.
We present our conclusions in section 4.

\section{\bf Weinberg-Salam Lagrangian}

Let us consider the bosonic sector of the Weinberg-Salam theory
\begin{equation}
{\cal L} = -\frac{1}{4} F_{\mu\nu}^a F^{\mu\nu,a}
- {1 \over 4} f_{\mu \nu} f^{\mu \nu}
+ (D_\mu \Phi)^{\dagger} (D^\mu \Phi)
- \lambda (\Phi^{\dagger} \Phi - \frac{v^2}{2} )^2
\   \end{equation}
with the SU(2)$_{\rm L}$ field strength tensor
\begin{equation}
F_{\mu\nu}^a=\partial_\mu V_\nu^a-\partial_\nu V_\mu^a
            + g \epsilon^{abc} V_\mu^b V_\nu^c
\ , \end{equation}
with the U(1) field strength tensor
\begin{equation}
f_{\mu\nu}=\partial_\mu A_\nu-\partial_\nu A_\mu
\ , \end{equation}
and the covariant derivative for the Higgs field
\begin{equation}
D_{\mu} \Phi = \Bigl(\partial_{\mu}
             -\frac{i}{2}g \tau^a V_{\mu}^a
             -\frac{i}{2}g' A_{\mu} \Bigr)\Phi
\ . \end{equation}
The gauge symmetry is spontaneously broken
due to the non-vanishing vacuum expectation
value $v$ of the Higgs field
\begin{equation}
    \langle \Phi \rangle = \frac{v}{\sqrt2}
    \left( \begin{array}{c} 0\\1  \end{array} \right)
\ , \end{equation}
leading to the boson masses
\begin{equation}
    M_W = \frac{1}{2} g v \ , \ \ \ \
    M_Z = {1\over2} \sqrt{(g^2+g'^2)} v \ , \ \ \ \
    M_H = v \sqrt{2 \lambda}
\ . \end{equation}
The mixing angle $\theta_w$ is determined by
the relation $ \tan \theta_w = g'/g $,
and the electric charge is $e = g \sin \theta_w$.

SU(2) gauge field configurations can be classified by a charge,
the Chern-Simons charge.
The Chern-Simons current
\begin{equation}
 K_\mu=\frac{g^2}{16\pi^2}\varepsilon_{\mu\nu\rho\sigma} {\rm Tr}(
 {\cal F}^{\nu\rho}
 {\cal V}^\sigma
 + \frac{2}{3} i g {\cal V}^\nu {\cal V}^\rho {\cal V}^\sigma )
\   \end{equation}
(${\cal F}_{\nu\rho} = 1/2 \tau^i F^i_{\nu\rho}$,
${\cal V}_\sigma = 1/2 \tau^i V^i_\sigma$)
is not conserved,
its divergence $\partial^\mu K_\mu$
represents the SU(2) part of the U(1) anomaly
of the baryon current.
The Chern-Simons charge
of a configuration is given by
\begin{equation}
N_{\rm CS} = \int d^3r K^0
\ . \end{equation}
For the vacua the Chern-Simons charge is identical to the
integer winding number,
while the sphaleron has a Chern-Simons charge of 1/2 [5].

Let us now consider the ansatz for the multisphaleron solutions.
Following Manton[13] and Rebbi and Rossi [14] we define
a set of orthonormal vectors
\begin{eqnarray}
\vec u_1^{(n)}(\phi) & = & (\cos n \phi, \sin n \phi, 0) \ ,
\nonumber \\
\vec u_2^{(n)}(\phi) & = & (0, 0, 1) \ ,
\nonumber \\
\vec u_3^{(n)}(\phi) & = & (\sin n \phi, - \cos n \phi, 0)
\ , \end{eqnarray}
and expand the fields as follows
\begin{equation}
V_i^a(\vec r) = u_j^{i(1)}(\phi) u_k^{a(n)}(\phi) w_j^k(\rho,z)
\ , \end{equation}
\begin{equation}
A_i(\vec r) = u_j^{i(1)}(\phi) a_j(\rho,z)
\ , \end{equation}
\begin{equation}
\Phi(\vec r) =i \tau^i u_j^{i(n)}(\phi) h_j(\rho,z)
      \frac{v}{\sqrt2}
    \left( \begin{array}{c} 0\\1  \end{array} \right)
\ . \end{equation}
Invariance under rotations about the $z$-axis
and under parity leads to the conditions [11,12]
\begin{equation}
w_1^1(\rho,z)=w_2^1(\rho,z)=w_1^2(\rho,z)=
		   w_2^2(\rho,z)=w_3^3(\rho,z)=0
\ , \end{equation}
\begin{equation}
a_1(\rho,z)=a_2(\rho,z)=0
\ , \end{equation}
\begin{equation}
h_3(\rho,z)=0
\ . \end{equation}

The resulting axially symmetric energy functional $E$
\begin{equation}
E = \frac{1}{2} \int ( E_w + E_a + v^2 E_h )
	      \, d\phi \, \rho d\rho \, dz
\   \end{equation}
then has the contributions
\begin{eqnarray}
E_w & = & (\partial_\rho w_3^1 + {1\over{\rho}} ( n w_1^3 + w_3^1 )
	 - g w_1^3 w_3^2 )^2
      +  (\partial_z    w_3^1 + {n\over{\rho}}   w_2^3
	 - g w_2^3 w_3^2 )^2
\nonumber \\
    & + &(\partial_\rho w_3^2 + {1\over{\rho}}   w_3^2
	 + g w_1^3 w_3^1 )^2
      +  (\partial_z    w_3^2
	 + g w_2^3 w_3^1 )^2
      +  (\partial_\rho w_2^3 - \partial_z w_1^3 )^2
\ , \end{eqnarray}
\begin{equation}
E_a = (\partial_\rho a_3 + {1\over{\rho}} a_3 )^2
       + (\partial_z a_3 )^2
\ , \end{equation}
\begin{eqnarray}
E_h & = & (\partial_\rho h_1 - {g\over2} w_1^3 h_2 )^2
       + (\partial_z    h_1 - {g\over2} w_2^3 h_2 )^2
       + (\partial_\rho h_2 + {g\over2} w_1^3 h_1 )^2
       + (\partial_z    h_2 + {g\over2} w_2^3 h_1 )^2
\nonumber \\
     & +& ({n\over{\rho}} h_1
       + {g\over2} ( w_3^1 h_2 - w_3^2 h_1 )
       - {g'\over2} a_3 h_1 )^2
       + (
         {g\over2} ( w_3^1 h_1 + w_3^2 h_2 )
       - {g'\over2} a_3 h_2 )^2
\nonumber \\
     & +& { {\lambda v^2}\over2} ( h_1^2 + h_2^2 - 1 )^2
\ . \end{eqnarray}
It is still invariant under gauge transformations generated by
\begin{equation}
 U= e^{i\Gamma(\rho,z) \tau^i u_3^{i(n)}}
\   \end{equation}
analogous to [11-14].
Note, that under this transformation the 2-D Higgs doublets
$(h_1,h_2)$ and $(w_3^1,w_3^2-n/g\rho)$ transform with
angle $\Gamma(\rho,z)$ and $2 \Gamma(\rho,z)$, respectively,
while the 2-D gauge field $(w_1^3,w_2^3)$ transforms
inhomogeneously.
Here we fix this gauge degree of freedom by choosing the
gauge condition [11,12]
\begin{equation}
\partial_\rho w_1^3 + \partial_z w_2^3 =0
\ . \end{equation}

Changing to spherical coordinates
and extracting the trivial $\theta$-dependence
we specify the ansatz further [11,12]
\nonumber \\
\begin{eqnarray}
w_1^3(r,\theta) \  &
= &  \ \ {2 \over{gr}} F_1(r,\theta) \cos \theta \ , \ \ \ \
w_2^3(r,\theta) \
= - {2 \over{gr}} F_2(r,\theta) \sin \theta    \ ,
\nonumber \\
w_3^1(r,\theta) \  &
= & - {{2 n}\over{gr}} F_3(r,\theta) \cos \theta    \ , \ \ \ \
w_3^2(r,\theta) \
=  \ \ {{2 n}\over{gr}} F_4(r,\theta) \sin \theta
\ , \end{eqnarray}
\begin{equation}
a_3(r,\theta) = {2 \over{g'r}} F_7(r,\theta) \sin \theta
\ , \end{equation}
\begin{equation}
h_1(r,\theta)   = F_5(r,\theta) \sin \theta \ , \ \ \ \
h_2(r,\theta)   = F_6(r,\theta) \cos \theta
\ . \end{equation}
Note, that the spherically symmetric ansatz for the sphaleron
corresponds to $n=1$ and
$F_1(r,\theta)=F_2(r,\theta)=F_3(r,\theta)=F_4(r,\theta)=f(r)$,
$F_5(r,\theta)=F_6(r,\theta)=h(r)$,
and $F_7(r,\theta)=0$
(where the functions
$f(r)$ and $h(r)$ correspond to those of ref. [5]).

The above ansatz, when inserted into the classical
equations of motion in the chosen gauge, yields a set of
coupled partial differential equations
for the functions $F_i(r,\theta)$, to be solved numerically
subject to certain boundary conditions.
To obtain regular, finite energy solutions
with the imposed symmetries, we take as
boundary conditions for the functions $F_i(r,\theta)$
\begin{eqnarray}
r=0 & :          & \ \ F_i(r,\theta)|_{r=0}=0,
               \ \ \ \ \ \ i=1,...,7 \ ,
\nonumber \\
r\rightarrow\infty& :
	       & \ \ F_i(r,\theta)|_{r=\infty}=1,
               \ \ \ \ \ i=1,...,6,
               \ \ \ F_7(r,\theta)|_{r=\infty}=0 \ ,
\nonumber \\
\theta=0& :      & \ \ \partial_\theta F_i(r,\theta)|_{\theta=0} =0,
	       \ \ \ i=1,...,7 \ ,
\nonumber \\
\theta=\pi/2& :  & \ \ \partial_\theta F_i(r,\theta)|_{\theta=\pi/2} =0,
               \ i=1,...,7
\ . \end{eqnarray}

The Chern-Simons charge of the multisphaleron solutions can be evaluated
analogously to the Chern-Simons charge of the sphaleron [5].
At spatial infinity
the vector fields are pure gauge configurations
\begin{equation}
\tau^a V_i^a = -\frac{2i}{g} \partial_i U U^\dagger
\   \end{equation}
with
\begin{equation}
U(\vec r \,) = \exp \Biggl( i \Omega(r,\theta)
  (\sin \theta \tau^i u_1^{i(n)} + \cos \theta \tau^i u_2^{i(n)}) \Biggr )
\ . \end{equation}
The proper gauge for evaluating the Chern-Simons charge is
the gauge with $U=1$ at infinity, which is obtained
with the transformation $U^\dagger$
and $\Omega(0)=0$ and $\Omega(\infty)=\pi/2$.
The Chern-Simons charge of the multisphaleron solutions
is then given by
\begin{equation}
N_{\rm CS} = {1 \over 2 \pi^2} \int d^3 r Q(\vec r \,)
\ , \end{equation}
where the Chern-Simons density is determined by [15,16]
\begin{equation}
Q(\vec r\,) =n
    { \sin^2 \Omega \over r^2} {\partial \Omega \over \partial r}
\ + \ {\rm derivative \ terms}
\ . \end{equation}
Since the derivative terms do not contribute (because of the
boundary conditions)
the Chern-Simons charge
of the multisphaleron solutions is given by
\begin{equation}
N_{\rm CS}= n/2
\ . \end{equation}
The Chern-Simons charge of the multisphalerons corresponds to their
baryonic charge, $Q_B = n/2$, since the U(1) field does not
contribute to their baryon number [5].

\section{\bf Multisphaleron solutions}

We here restrict the numerical calculations
to the limit of vanishing mixing angle.
In this limit the U(1) field
decouples and the function $F_7(r,\theta)$
can consistently be set to zero.
The case of non-vanishing mixing angle will be considered later [15].
Subject to the boundary conditions (25) we then solve the equations
numerically, using the dimensionless coordinate $x=gvr$.
We fix the parameters $g=0.65$ and $M_W=80 {\rm GeV}$.
The Higgs mass is varied.

The numerical calculations are based on the Newton-Raphson
method. The equations are discretized on a non-equidistant
grid in $x$ and an equidistant grid in $\theta$,  where
typical grids used have sizes $50 \times 20$ and $100 \times 20$
covering integration regions $0<x<60$ up to $0<x<180$ and
$0<\theta<\pi/2$.
The numerical error for the functions is estimated to be
on the order of $10^{-3}$.

To obtain the multisphaleron solutions we start with the $n=1$
sphaleron solution as initial guess and increase the value of $n$
slowly. This numerical procedure thus involves the construction
of intermediate configurations, where
$n$ takes on non-integer values.
At each integer values of $n$
a multisphaleron solution is reached.

In the following we exhibit
the axially symmetric multisphaleron solutions.
In particular we show as gauge invariant quantities,
the energy density $\varepsilon$,
defined by
\begin{equation}
E= \frac{1}{4\pi} \int \varepsilon (\vec x)
                       x^2 dx \sin \theta d\theta d\phi
\ , \end{equation}
where $E$ is the energy in TeV,
and the magnitude of the Higgs field $\Phi$ in units of $v/\sqrt{2}$.
In Figs.~1a and 1b we show the energy density $\varepsilon$
of the multisphalerons
with $n=2$ and $n=5$ for the Higgs mass $M_H=M_W$,
and in Figs.~2a and 2b we show the corresponding
magnitude of the Higgs field $\Phi$.
The energy density is strongly peaked along the $\rho$-axis,
with the maximum shifting outward with increasing $n$.
The magnitude of the Higgs field, being zero at the origin,
remains small along the $\rho$-axis for increasingly longer
intervals with increasing $n$.
This is also demonstrated in the following comparison
of the sphaleron and all multisphalerons up to $n=5$.
In Figs.~3a and 3b we show
the energy density along the $\rho$-axis and along the $z$-axis
for the multisphalerons up to $n=5$ for $M_H=M_W$,
and in Figs.~4a and 4b we show the corresponding
magnitude of the Higgs field.

The energy $E(n)$ of the multisphalerons from $n=2$ to $n=5$
is shown in Table 1 for the Higgs masses
$M_H= M_W/10$, $M_H=M_W$ and $M_H= 10 M_W$
and compared with the corresponding energy $E(1)$ of the sphaleron,
by considering the ratio $E(n)/nE(1)$.
While for small Higgs masses the energy of the multisphalerons is smaller
than $n$ times the energy of the sphaleron,
for large Higgs masses
the energy of the multisphalerons is larger
than $n$ times the energy of the sphaleron.

\section{\bf Conclusions}

We have constructed multisphaleron solutions in the weak interactions
for vanishing mixing angle.
The extension to finite mixing angle is in progress [15].

While the sphaleron is spherically symmetric for vanishing
mixing angle,
the multisphaleron solutions are only axially symmetric.
They can be characterized by an integer $n$,
related to the winding of the fields in the azimuthal angle $\phi$.
With increasing $n$ the energy density of the
multisphalerons becomes increasingly deformed,
the maximum occurring along the $\rho$-axis
at increasing distance from the origin.

The energy of the multisphalerons
is on the order of $n$ times the sphaleron energy.
For small Higgs masses the energy
of the multisphalerons is less than $n$ times
the energy of the sphaleron, $E(n)< nE(1)$,
while for large Higgs masses the energy
of the multisphalerons is greater than $n$ times
the energy of the sphaleron, $E(n)> nE(1)$.

While the sphaleron has Chern-Simons charge $N_{\rm CS}=1/2$,
the multisphaleron solutions have
Chern-Simons charge $N_{\rm CS}=n/2$.
We conjecture, that symmetric
non-contractible loops can be constructed,
leading from a vacuum with Chern-Simons charge $N_{\rm CS}=0$
to a topologically distinct vacuum with
Chern-Simons charge $N_{\rm CS}=n$,
passing the multisphaleron solution with $N_{\rm CS}=n/2$
midway.

At finite temperature such paths involving multisphalerons
should allow for thermal
fermion number violating transitions,
corresponding to
tunnelling transitions via multiinstantons
at zero temperature.
Since multiinstantons with winding number $n$
possess $n$ fermion zero modes,
we expect to encounter $n$ fermion zero modes
along vacuum to vacuum paths, passing the $n$-th multisphaleron.

{\bf \sl Acknowledgement}

We gratefully acknowledge discussions with Yves Brihaye.

\vfill\eject

\vfill\eject
\section{\bf Table}
\begin{center}
Table 1 \\
\vspace{3 mm}
$ E(n)/{\rm TeV}$   $(E(n)/nE(1)) $ \\
\vspace{3 mm}
\begin{tabular}{||c|c|c|c||} \hline\hline
$n$   & $M_H =  M_W/10 $ & $ M_H = M_W$ & $M_H = 10 M_W $\\
\hline
$ 1 $ & $ 7.468$ $(1.000)$ & $ 8.665$ $(1.000)$  & $11.375$ $(1.000)$ \\
$ 2 $ & $13.791$ $(0.923)$ & $17.140$ $(0.989)$  & $24.937$ $(1.096)$ \\
$ 3 $ & $19.881$ $(0.887)$ & $25.879$ $(0.996)$  & $40.030$ $(1.173)$ \\
$ 4 $ & $25.908$ $(0.867)$ & $34.931$ $(1.008)$  & $56.298$ $(1.237)$ \\
$ 5 $ & $31.931$ $(0.855)$ & $44.288$ $(1.022)$  & $73.521$ $(1.293)$ \\
\hline
\hline
\end{tabular}
\end{center}

\section{\bf Figure captions}

{\bf Figure 1:}

The energy density $\varepsilon$ (in TeV) is shown
as a function of the dimensionless coordinates $\rho$ and $z$
for the multisphalerons
with $n=2$ (1a) and $n=5$ (1b).

{\bf Figure 2:}

The magnitude of the Higgs field $\Phi$ (in units of $v/\sqrt{2}$)
is shown
as a function of the dimensionless coordinates $\rho$ and $z$
for the multisphalerons
with $n=2$ (2a) and $n=5$ (2b).

{\bf Figure 3:}

The energy density $\varepsilon$ (in TeV) along the $\rho$-axis (3a)
and $z$-axis (3b) is shown
as a function of the dimensionless coordinate $\rho$ resp.\ $z$
for the sphaleron with $n=1$ and the multisphalerons
from $n=2$ to $n=5$.

{\bf Figure 4:}

The magnitude of the Higgs field $\Phi$ (in units of $v/\sqrt{2}$)
along the $\rho$-axis (3a)
and $z$-axis (3b) is shown
as a function of the dimensionless coordinate $\rho$ resp.\ $z$
for the sphaleron with $n=1$ and the multisphalerons
from $n=2$ to $n=5$.


\begin{thebibliography}{000}
\bibitem{1.}
G. 't Hooft,
Symmetry breaking through Bell-Jackiw Anomalies,
Phys. Rev. Lett. 37 (1976) 8.

\bibitem{2.}
A. Ringwald,
High energy breakdown of perturbation theory in the electroweak
instanton sector,
Nucl. Phys. B330 (1990) 1.

\bibitem{3.}
M. Mattis, and E. Mottola, eds.,
``Baryon Number Violation at the SSC?'',
World Scientific, Singapore (1990).

\bibitem{4.}
N.~S. Manton,
Topology in the Weinberg-Salam theory,
Phys. Rev. D28 (1983) 2019.

\bibitem{5.}
F.~R. Klinkhamer, and N.~S. Manton,
A saddle-point solution in the Weinberg-Salam theory,
Phys. Rev. D30 (1984) 2212.

\bibitem{6.}
V.~A. Kuzmin, V.~A. Rubakov, and M.~E. Shaposhnikov,
On anomalous electroweak baryon-number non-conservation
in the early universe,
Phys. Lett. B155 (1985) 36.

\bibitem{7.}
P. Arnold, and L. McLerran,
Sphalerons, small fluctuations, and baryon-number violation
in electroweak theory,
Phys. Rev. D36 (1987) 581.

\bibitem{8.}
P. Arnold, and L. McLerran,
The sphaleron strikes back: A response
to objections to the sphaleron approximation,
Phys. Rev. D37 (1988) 1020.

\bibitem{9.}
L. Carson, X. Li, L. McLerran, and R.-T. Wang,
Exact computation of the small-fluctuation determinant
around a sphaleron,
Phys. Rev. D42 (1990) 2127.

\bibitem{10.}
E.~W. Kolb, and M.~S. Turner,
``The Early Universe'',
Addison-Wesley Publishing Company, Redwood City (1990).

\bibitem{11.}
B. Kleihaus, J. Kunz, and Y. Brihaye,
The electroweak sphaleron at physical mixing angle,
Phys. Lett. B273 (1991) 100.

\bibitem{12.}
J. Kunz, B. Kleihaus, and Y. Brihaye,
Sphalerons at finite mixing angle,
Phys. Rev. D46 (1992) 3587.

\bibitem{13.}
N. S. Manton,
Complex structure of monopoles,
Nucl. Phys. B135 (1978) 319.

\bibitem{14.}
C. Rebbi and P. Rossi,
Multimonopole solutions in the Prasad-Sommerfield limit,
Phys. Rev. D22 (1980) 2010.

\bibitem{15.}
B. Kleihaus and J. Kunz,
in preparation

\bibitem{16.}
Y. Brihaye and J. Kunz,
in preparation

\end{thebibliography}
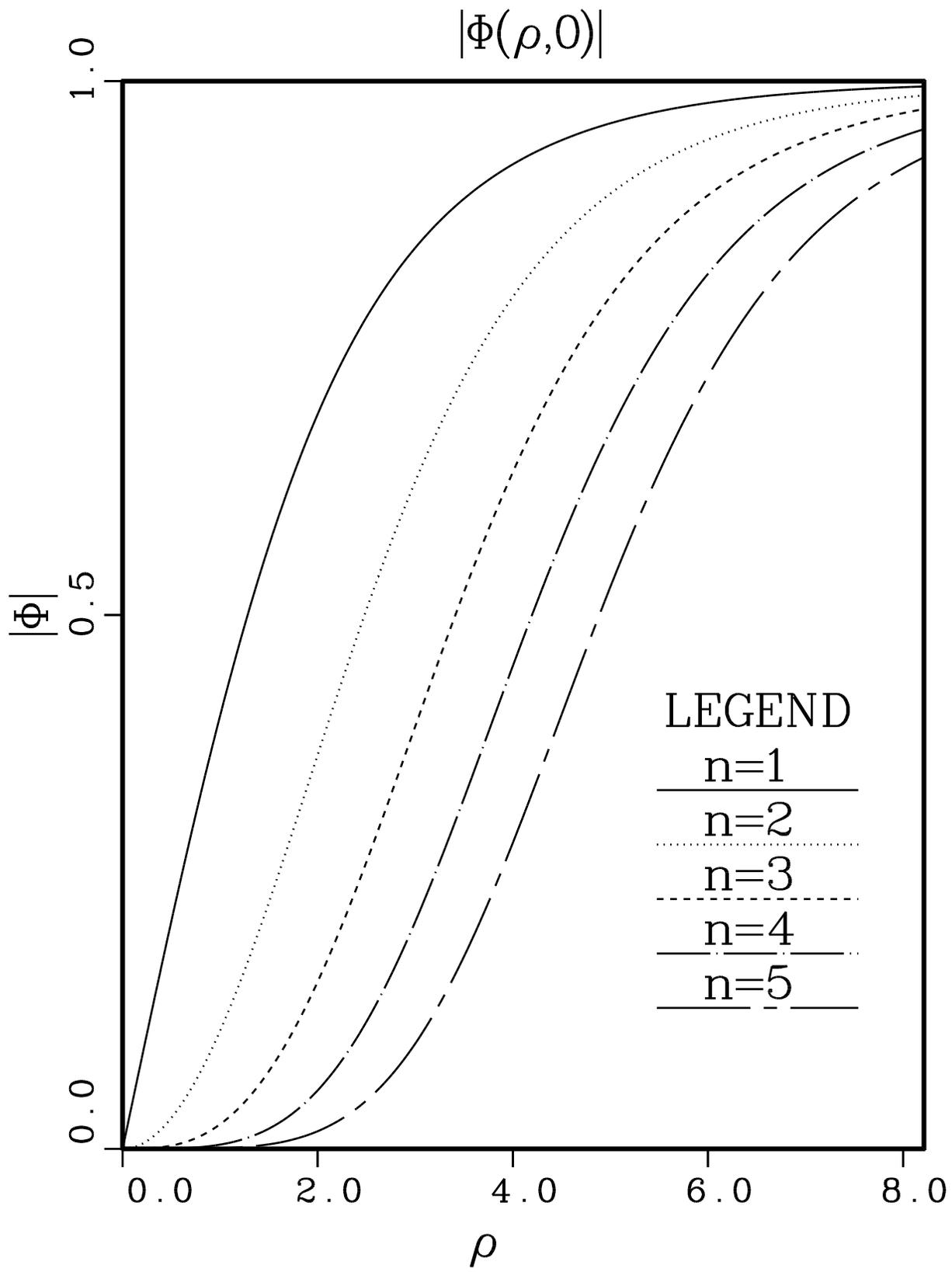
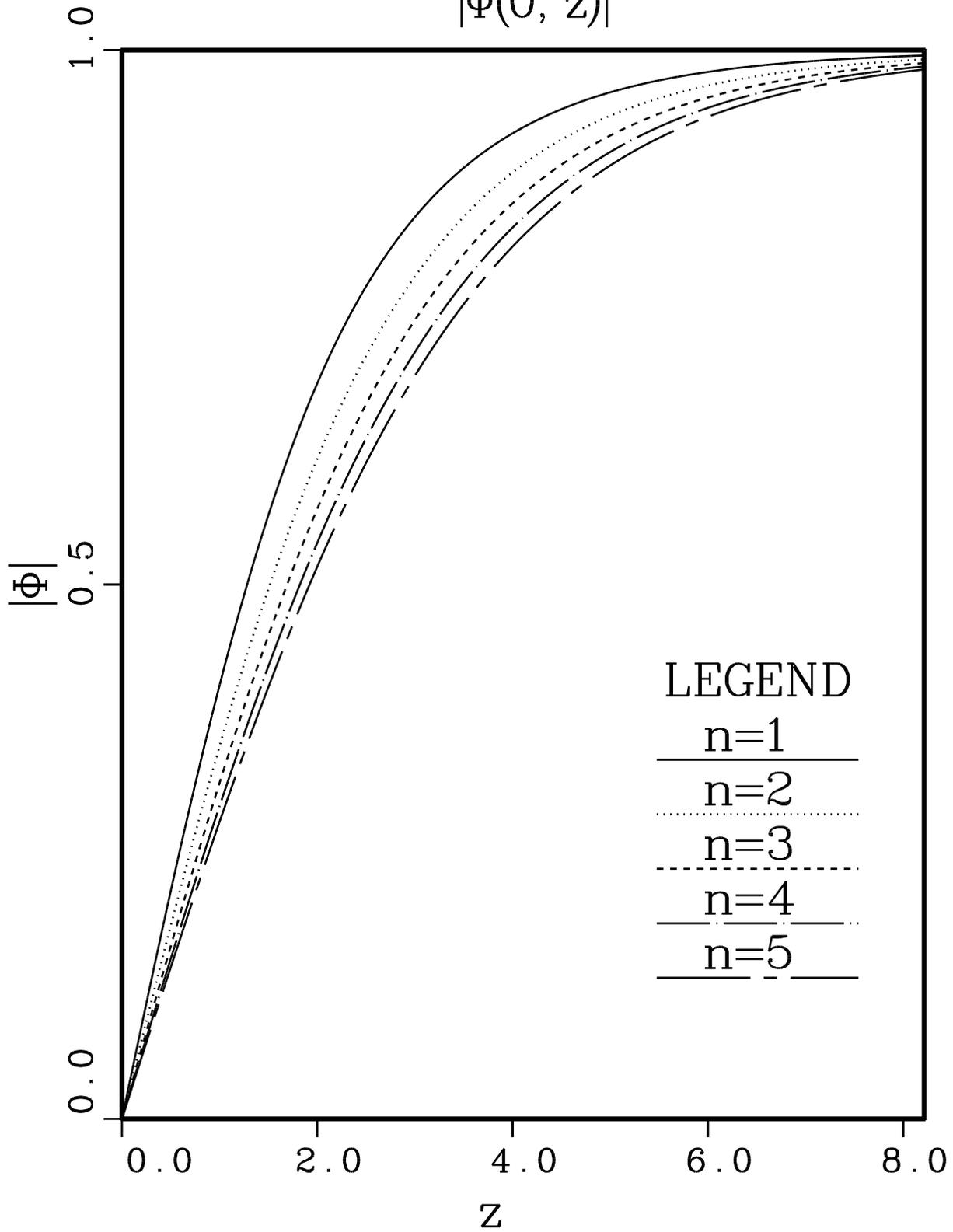
\end{document}